\documentclass[pre,reprint,amssymb,amsmath,amsfonts,aps,floatfix,showpacs]{revtex4-1}
\usepackage{graphicx, units, enumerate} 

\newcommand{\Epi}{\affiliation{Department of Epileptology, University of Bonn, \\Sigmund-Freud-Stra{\ss}e~25, 53105~Bonn, Germany}}
\newcommand{\HISKP}{\affiliation{Helmholtz Institute for Radiation and Nuclear Physics, University of Bonn, \\Nussallee~14--16, 53115~Bonn, Germany}}
\newcommand{\IZKS}{\affiliation {Interdisciplinary Center for Complex Systems, University of Bonn, \\Br{\"u}hler Stra{\ss}e~7, 53175~Bonn, Germany}}

\newcommand{\kl}[1]{\left( #1 \right)}
\newcommand{\klg}[1]{\left\{ #1 \right\}}
\newcommand{\kle}[1]{\left[ #1 \right]}

\newcommand{\ed}[1]{\frac{1}{#1}}
\newcommand{\defi}{\mathrel{\mathop:}=}
\newcommand{\ifed}{=\mathrel{\mathop:}}

\begin{document}

\title{Constrained randomisation of weighted networks}

\author{Gerrit Ansmann}
\email{gansmann@uni-bonn.de}
\author{Klaus Lehnertz}
\email{klaus.lehnertz@ukb.uni-bonn.de}
\Epi \HISKP \IZKS

\begin{abstract}
We propose a Markov chain method to efficiently generate \textit{surrogate networks} that are random under the constraint of given vertex strengths.
With these strength-preserving surrogates and with edge-weight-preserving surrogates we investigate the clustering coefficient and the average shortest path length of functional networks of the human brain as well as of the International Trade Networks.
We demonstrate that surrogate networks can provide additional information about network-specific characteristics and thus help interpreting empirical weighted networks.
\end{abstract}


\maketitle
\section{Introduction}
Over the past decade, network theory has contributed significantly to improve our understanding of collective dynamics in networks with complex topologies.
The simplicity of the network representation, where the interactions and interacting elements are mapped to edges and vertices, respectively, stimulated its use on a number of systems, ranging from physical, biological to social and engineering systems \cite{Strogatz2001, Albert2002, Newman2003, Boccaletti2006a, Sporns2006, BarratBook2008, Reijneveld2007, Arenas2008, Bullmore2009, Donges2009b, Tsonis2010, Qiu2010}.
A large number of natural and man-made systems have been shown to be neither entirely regular nor entirely random, but to exhibit prominent topological properties, such as short average path lengths and a high level of clustering.

Recently, weighted networks, in which each edge is assigned a weight, have been shown to allow a better description of many natural and man-made systems \cite{Yook2001, Barrat2004b, Newman2004, Chavez2005, Boccaletti2006a, Onnela2007, Arenas2008}, and particularly of functional networks underlying various brain pathologies \cite{Rubinov2009, Stam2009, Ponten2009, Chavez2010, Horstmann2010, Wang2010b}.
Functional brain networks are usually derived from either direct or indirect measurements of neural activity.
Network vertices are associated with sensors that are placed such as to sufficiently capture the dynamics of different brain regions.
The connectedness between any pair of brain regions is assessed by evaluating some linear or non-linear interdependencies between their neural activities \cite{Pikovsky_Book2001, Pereda2005, Hlavackova2007, Lehnertz2009b}.
Such networks can be regarded as complete weighted networks, in which all possible edges exist.

For empirical networks, interpreting findings is not without challenges.
Findings of some network characteristics may be influenced by statistical fluctuations (like measurement or environmental noise) and systematic errors (which might, for example, be attributed to the data acquisition or to the selected way to construct a network from the data).
Moreover, existing methods of analysis may be misapplied or misinterpreted, which may lead to inappropriate conclusions, as pointed out in Refs.~\cite{Butts2009, Bialonski2010, vanWijk2010}.
Standard approaches to uncovering influencing factors like background measurements, repeated measurements, or selective manipulation of the investigated system may, however, not be feasible in empirical network studies.
Another strategy is the comparison with the expected result for appropriate null models.
This result can either be derived analytically \cite{Newman2001b, Foster2007, Garlaschelli2009} or be extracted from samples that are obtained by Monte Carlo simulations \cite{Maslov2002, Maslov2004, Barrat2004b, Sporns2004, Randrup2005, Serrano2006, Serrano2008, Opsahl2008, Zlatic2009, DelGenio2010}.
In the following we refer to these samples as `surrogate networks', in accordance with a similar approach, that is well established in time series analysis \cite{Theiler1992, Schreiber2000a}.

We here propose an efficient iterative procedure to generate strength-preserving surrogate networks for investigations of complete weighted networks.
This paper is organised as follows.
In Sec.~\ref{methods} we describe our approach to surrogate networks and introduce our procedure.
We show that it generates approximately uniformly distributed surrogates for a sufficient number of iterations and propose a method to determine this number.
With strength-preserving surrogates and weight-preserving surrogates we reanalyse functional networks of the human brain and investigate the International Trade Networks (Sec.~\ref{appl}).
We demonstrate that surrogates can provide additional information about network-specific characteristics and thus aid in their interpretation.
Finally, in Sec.~\ref{concl} we draw our conclusions.

\section{Methods}
\label{methods}
\subsection{Definitions and Measures}
\label{defmeas}
We consider undirected, weighted networks with non-negative edge weights and treat them as complete networks, i.e., we consider every possible edge to exist.
A network of this type with $n$ vertices is fully described by its symmetric non-negative weight matrix $W \in \mathbb{R}_+^{n \times n}$, whose entry $W_{ij}$ is the weight of the edge connecting vertices $i$ and $j$.
For practical purposes we define the diagonal elements $W_{ii}$ as zero.
The strength of a vertex is defined as the sum of all adjacent weights $S_i \defi \sum_{j=1}^n W_{ij}$.
We consider the distribution of all edge weights of a network $\mathcal{W} \defi \klg{W_{12}, W_{13}, W_{23}, \ldots, W_{n-1,n}}$ and the distribution of all vertex strengths $\mathcal{S} \defi \klg{S_1, \ldots S_n}$.

For the weighted clustering coefficient of node $i$ we use the following definition \cite{Onnela2005}:
\[C_i \defi \frac{\sum\limits_{jk} \sqrt[3]{W_{ij} W_{jk} W_{ki}}} { \kl{n-1} \kl{n-2} \max\kl{\mathcal{W}} }.\]
This definition has the advantage that the value of the clustering coefficient is continuous for $W_{ij} \rightarrow 0$ \cite{Saramaki2007}.
We also consider
\[K_i \defi \frac{\sum\limits_{jk} \sqrt[3]{W_{ij} W_{jk} W_{ki}}} { \kl{n-1} \kl{n-2}} = C_i \max\kl{\mathcal{W}}.\]

For the weighted shortest path $L_{ij}$ between vertices $i$ and $j$ we follow Ref.~\cite{Newman2001c}
and consider the inverse of the weight of an edge as the length of that edge.

As network specific characteristics we here investigate the averages $\bar{C}$, $\bar{K}$, and $\bar{L}$ of $C_i$, $K_i$, and $L_{ij}$, respectively.

\subsection{Network Surrogates}
We consider the extent, to which distributions of local network properties (such as $\mathcal{W}$ or $\mathcal{S}$) contribute to the network-specific characteristic under investigation (such as $\bar{C}$, $\bar{K}$, or $\bar{L}$). In many situations this quantity may reveal important aspects of the network or of the applied methods:
\begin{itemize}
	\item If the edge weights---instead of being determined by the investigated system---are independently drawn from some distribution (e.g., due to excessive noise), the value of any characteristic can only be attributed to the weight distribution $\mathcal{W}$ and to coincidence.
	\item Edge weights defined from the data are often normalised by multiplication with a factor, that depends on a distribution of local properties (e.g., the average strength $\bar{\mathcal{S}}$).
	This usually changes the extent, to which this distribution contributes to network-specific characteristics.
	Sign and magnitude of this change may help to decide, whether a normalisation works as intended.
	\item If the weight of an edge only depends monotonically on some intrinsic property of its adjacent vertices (e.g., in fitness model networks \cite{Caldarelli2002, Barrat2004}), the value of network-specific characteristics may be mainly attributed to the strength distribution $\mathcal{S}$.
	\item If one local entity (e.g., an edge weight) dramatically exceeds the others in some local property (e.g., if the maximum edge weight is by far larger than the other weights), it may dominate a network-specific characteristic.
	As this influence is mediated by the distribution of this property, the network-specific characteristic would be mainly attributed to this distribution.
	\item If the value of a network-specific characteristic can be fully attributed to the distribution of a local network property, it should be considered, whether in this case a network approach to the data is overly complicated and more simple properties may be regarded instead.
\end{itemize}

To decide, to which extent a characteristic of a given network (the `original network') is determined by the distribution of a local network property, it can be compared to the values for surrogates of this network, which are randomised under the constraint that this distribution is preserved.
Moreover, the null hypothesis can be tested, that the 
network under consideration is random under the constraint of the distribution of the local property.
Details about null hypothesis tests based on surrogates can be found in the literature, e.g., in Ref.~\cite{Schreiber2000a}.

We here consider surrogate methods, which exactly preserve either the strength distribution $\mathcal{S}$ or the weight distribution $\mathcal{W}$ (preserving both would in most cases only leave one possible surrogate network, namely, the original network). We aim at methods that sample uniformly from the set of all networks with a given $\mathcal{S}$ or $\mathcal{W}$, respectively.
The corresponding null hypotheses are
\begin{description}
	\item[$H_\mathcal{S}$] The network under consideration is random under the constraint of its strength distribution $\mathcal{S}$.
	\item[$H_\mathcal{W}$] The network under consideration is random under the constraint of its weight distribution $\mathcal{W}$.
\end{description}
Note that preserving the strength distribution is equivalent to preserving the strength sequence when regarding network-specific properties, since they are not affected by a permutation of the vertices.
While the generation of uniformly-distributed weight-preserving surrogates can be achieved by a reshuffling of the weights \cite{Barrat2004b, Opsahl2008}, our method to generate strength-preserving surrogates is described in the following.

\subsection{Strength-preserving surrogate networks}
\label{sps}
The constraint of a given strength sequence of an undirected, weighted, and complete network with $n$ vertices can be expressed by a system of $n$ linear equations with the $m \defi \frac{1}{2}n\kl{n-1}$ edge weights as variables.
Given the non-negativity of the edge weights the set of solutions to this set of equations represents a convex polytope $\Omega \in \mathbb{R}^m$ \cite{Shephard1971}, each point of which corresponds to a network.
Thus the problem of generating strength-preserving surrogates is equivalent to that of picking random points from a polytope.
Some exact solutions to this problem (e.g., utilizing triangulation) have been proposed \cite{Rubin1984}, but due to computational burden they may be applied to networks with a very small number of vertices only.
Hit-and-Run samplers \cite{Smith1996} are a group of iterative Monte-Carlo procedures providing samples from a bounded region, such as a polytope.
The distribution of these samples has been shown to approximate the uniform distribution on that region under certain requirements and for a sufficient number of iterations \cite{Smith1984}.
We here propose a Hit-and-Run sampler, that is specialised to the problem of generating strength-preserving surrogates.
In Appendix~\ref{mathematics} we present a mathematical background to this procedure and show, that it fulfils the requirements for sampling approximately uniform.

\subsubsection{Procedure}
\label{procedure}
\begin{figure}
	\includegraphics{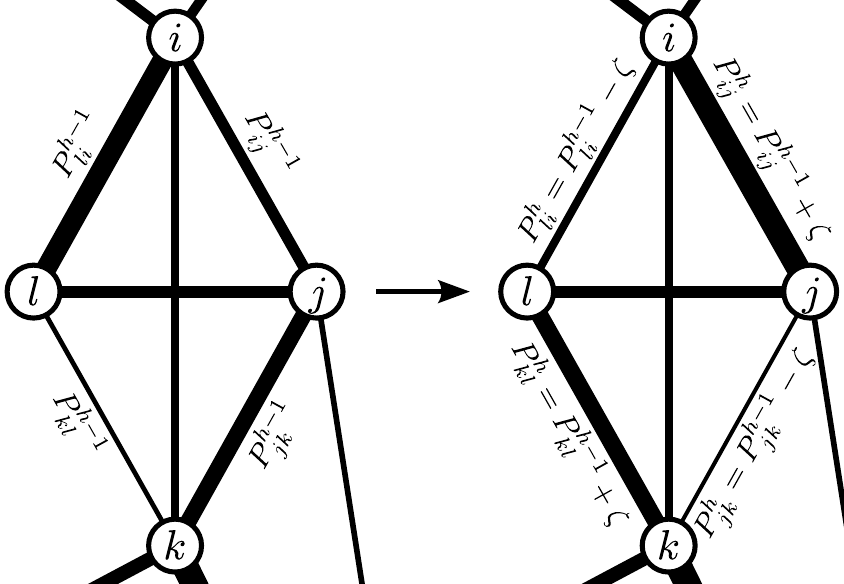}
	\caption{`Tetragon transformation' of the network $P^{h-1}$ to $P^{h}$.
	In a randomly selected tetragon ($i$, $j$, $k$, $l$) a random number $\zeta$ is added to the weights of two opposing edges ($P_{ij}^{h-1}$, $P_{kl}^{h-1}$) and subtracted from the weights of the others ($P_{jk}^{h-1}$, $P_{li}^{h-1}$).
	Other edge weights remain unaltered.
	Edge weights are encoded as line thickness.}
	\label{fig:Quadrate}
\end{figure}

We propose the following procedure for sampling from the set $\Omega$ of all networks with a given strength sequence:
\begin{enumerate}
	\item Acquire some network $P^0 \in \Omega$ and set the counter $h=1$.
	\item \label{algopick} Randomly select four pairwise distinct vertex indices $i,j,k,l \in \klg{1,\ldots,n}$.
	\item Pick a number $\zeta$ from the uniform distribution on $\kle{-\min\kl{P^{h-1}_{ij}, P^{h-1}_{kl}}, \min\kl{P^{h-1}_{jk}, P^{h-1}_{li}}}$.
	Let $P^h = P^{h-1}$, but set $P^h_{ij} = P^{h-1}_{ij} + \zeta$, $P^h_{jk} = P^{h-1}_{jk} - \zeta$, and $P^h_{kl} = P^{h-1}_{kl} + \zeta$, $P^h_{li} = P^{h-1}_{li} - \zeta$ (cf. Fig.~\ref{fig:Quadrate}).
	\item If $h<t$, raise $h$ by $1$ and continue at \ref{algopick}.
	Otherwise let $P^t$ be the surrogate network.
\end{enumerate}

The interval, to which $\zeta$ is limited, is the maximum one, such that the transformed network does not contain any negative weights.
This procedure can be regarded as an extension of previously suggested null model samplers \cite{Maslov2002, Maslov2004, Serrano2008, Zlatic2009}.

In principle, $P^t$ as generated by our procedure is statistically dependent on $P^0$.
This dependence becomes negligible, however, for a sufficiently high number of transformations $t_\text{suf}$ (to be determined in Sec.~\ref{Konvergenz}).
Most computational effort has to be spent reducing this statistical dependence.

\begin{figure}
	\includegraphics{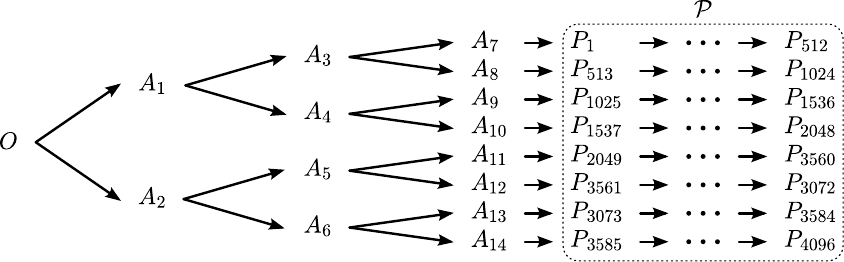}
	\caption{Scheme used to generate surrogates $\mathcal{P}=\klg{P_1, \ldots P_{4096}}$ from an original network $O$.
	Short arrows represent a `step' consisting of $t$ tetragon transformations, long arrows represent ten such steps, $A_1, \ldots, A_{14}$ are auxiliary networks.
	(Superscripts are omitted for better readability.)}
	\label{fig:kettenbaum}
\end{figure}

Concerning the acquisition of the starting point $P^0$, the most direct approach would be to select $P^0_i = O ~\forall i$, where $O$ is the original network and the subscript index here indicates different surrogates to be generated.
This way, however, the reduction of statistical dependence achieved when generating surrogate $P^t_{i-1}$ is discarded when generating surrogate $P^t_i$.
To benefit more from previously achieved reductions of dependence, we therefore employed schemes, where $P^0_i$ is a previously generated surrogate for most $i$ (e.g., $P^0_i=P^t_{i-1}$).
Out of several such schemes, the one depicted in Fig.~\ref{fig:kettenbaum} required the smallest number of total iterations to generate $4096$ surrogates with negligible dependencies (according to the test presented in Sec.~\ref{Konvergenz}).
This scheme was roughly ten times faster than the direct generation of surrogates from the original network ($P^0_i = O$).

\subsubsection{Numerical estimation of the necessary number of transformations}
\label{Konvergenz}
In order to estimate, which number $t$ of transformations is sufficient, we employed the following procedure to test whether surrogate networks are sampled appropriately.
It estimates the likelihood that surrogates $\mathcal{P} = \klg{P_1, \ldots, P_a}$ ($a \in \mathbb{N}$) are picked independently from the uniform distribution.
\begin{enumerate}
	\item Select parameters $b, c \in \mathbb{N}$ with $c \ll a$.
	\item Generate $a$ surrogates $\mathcal{Q} = \klg{Q_1, \ldots, Q_a}$ with a `reference method' that is known to pick surrogates independently from the uniform distribution.
	Pick some random testing points $\mathcal{R} = \klg{R_1, \ldots, R_b}$ from the polytope $\Omega$, e.g., by using the reference method.
	\item For all $i \in \klg{1, \ldots, b}$, determine $\epsilon_i > 0$ such that exactly $c$ surrogates from $\mathcal{Q}$ are in the $\epsilon_i$-ball around $R_i$.
	\item For all $i \in \klg{1, \ldots, b}$, let $k_i$ be the number of surrogates from $\mathcal{P}$ in the $\epsilon_i$-ball around $R_i$.
	\item Let $\tilde{p}\kl{k} \defi \binom{a}{k} \binom{2a}{c+k}^{-1}$ and $\chi \defi \ed{b} \kl{\sum\limits_{i=1}^{b} \tilde{p}\kl{k_i}} \kl{\sum\limits_{j=0}^{a} \tilde{p}\kl{j}}  \kl{\sum\limits_{j=0}^a \tilde{p}\kl{j}^2}^{-1}.$ The expected value of $\chi$ is $1$ if $P_1, \ldots, P_a$ are picked independently from the uniform distribution.
	Otherwise and if $a$ and $b$ are sufficiently high and $c$ is sufficiently low, the expected value of $\chi$ is lower than $1$ (cf. App.~\ref{appprob} for details).

\end{enumerate}

\begin{figure}
	\input{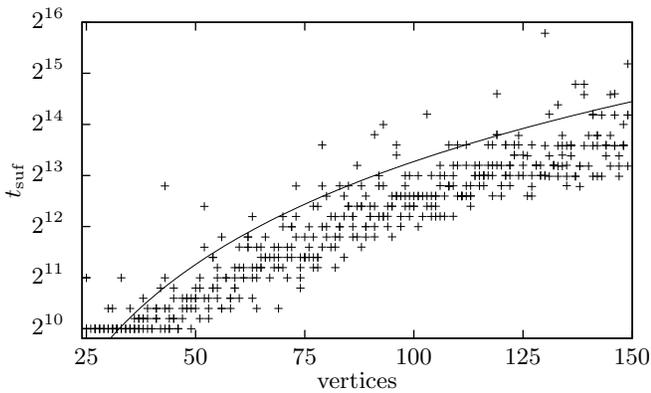}
	\caption{Number of sufficient transformations $t_\text{suf}$ per step for the generation of $4096$ surrogate networks from four toy networks with $n$ vertices.
	For comparison, the solid line displays twice the number of edges.}
	\label{fig:konvtest}
\end{figure}

To estimate the necessary number $t$ of transformations per step (cf. Fig.~\ref{fig:kettenbaum}), we regarded four toy networks with random weights for each number of vertices between $25$ and $149$.
We raised $t$ from $1024$ successively by a factor of $2^{0.2}$.
For each $t$ we generated several realisations $\mathcal{P}$ of $4096$ surrogates each and if $\chi > 0.96$ for each $\mathcal{P}$, we set $t_\text{suf}=t$.
As a reference method we used the same method with $t=2^{19}$, which we assumed to generate appropriately sampled surrogates.
To avoid the reference $\mathcal{Q}$ being statistically outlying, however, we omitted it, if it scored $\chi < 0.98$ in a test against another reference generated by the same method.
For comparison, $\chi=1.01 \pm 0.03$ for the reference methods in a test against themselves.
In Fig.~\ref{fig:konvtest} we show the number of sufficient transformations $t_\text{suf}$ for different numbers of vertices of the toy networks.
We observe that in most cases our method generates appropriate surrogates if $t$ is approximately twice the number of edges in the original network.

Generating $4096$ surrogate networks with $t=2^{16}$ transformations per step took \unit[123]{s} on a PC with \unit[829]{MFLOPS} (\unit[2]{GHz}).

\section{Surrogate analysis of empirical networks}
\label{appl}
\subsection{Functional brain networks}
\label{hirn}
Characterizing anatomical and functional connections in the human brain with approaches from network theory has been a rapidly evolving field recently \cite{Reijneveld2007, Arenas2008, Bullmore2009}.
Research over the past years indicates that both physiological and pathophysiological states of the brain are reflected by topological aspects of functional brain networks.
Mostly the clustering coefficient, the average shortest path length or similar measures had been used to characterise these networks.
Findings that had been achieved so far can be regarded as important since they provide new insights into properties of normal and pathologic functional brain networks.

In Ref.~\cite{Horstmann2010} functional brain networks derived from electroencephalographic (EEG) recordings during different states of vigilance (eyes opened and eyes closed) of 21 epilepsy patients and of 23 healthy control subjects had been analysed using the clustering coefficient $\bar{C}$ and the average shortest path length $\bar{L}$.
Differences in these characteristics could be observed between epilepsy patients and healthy control subjects as well as between states of vigilance.
We here reanalysis exemplary networks from an epilepsy patient and a healthy control subject, and with surrogate networks we investigated to which extent the observed findings can be attributed to the weight distribution $\mathcal{W}$ or strength distribution $\mathcal{S}$.

Details of the data and of recording and analysis techniques are fully described in Ref.~\cite{Horstmann2010}.
Briefly, EEG data had been recorded for \unit[30]{min} with $n=29$ electrodes
\footnote{Fp1, Fp2, F7, F3, Fz, F4, F8, FC1, FC2, T7, C3, Cz, C4, T8, CP1, CP2, P7, P3, Pz, P4, P8, PO7, PO3, PO4, PO8, Oz, O9, Iz, and O10}
placed according to the 10-10 system of the American Electroencephalographic Society with the right mastoid as physical reference (sampling rate: \unit[254.31]{Hz}; \unit[16]{bit} A/D conversion; bandwidth: \unit[0--50]{Hz}).
During one half of the recording time each subjects had their eyes opened or closed, respectively.

\begin{figure*}[p]
	\input{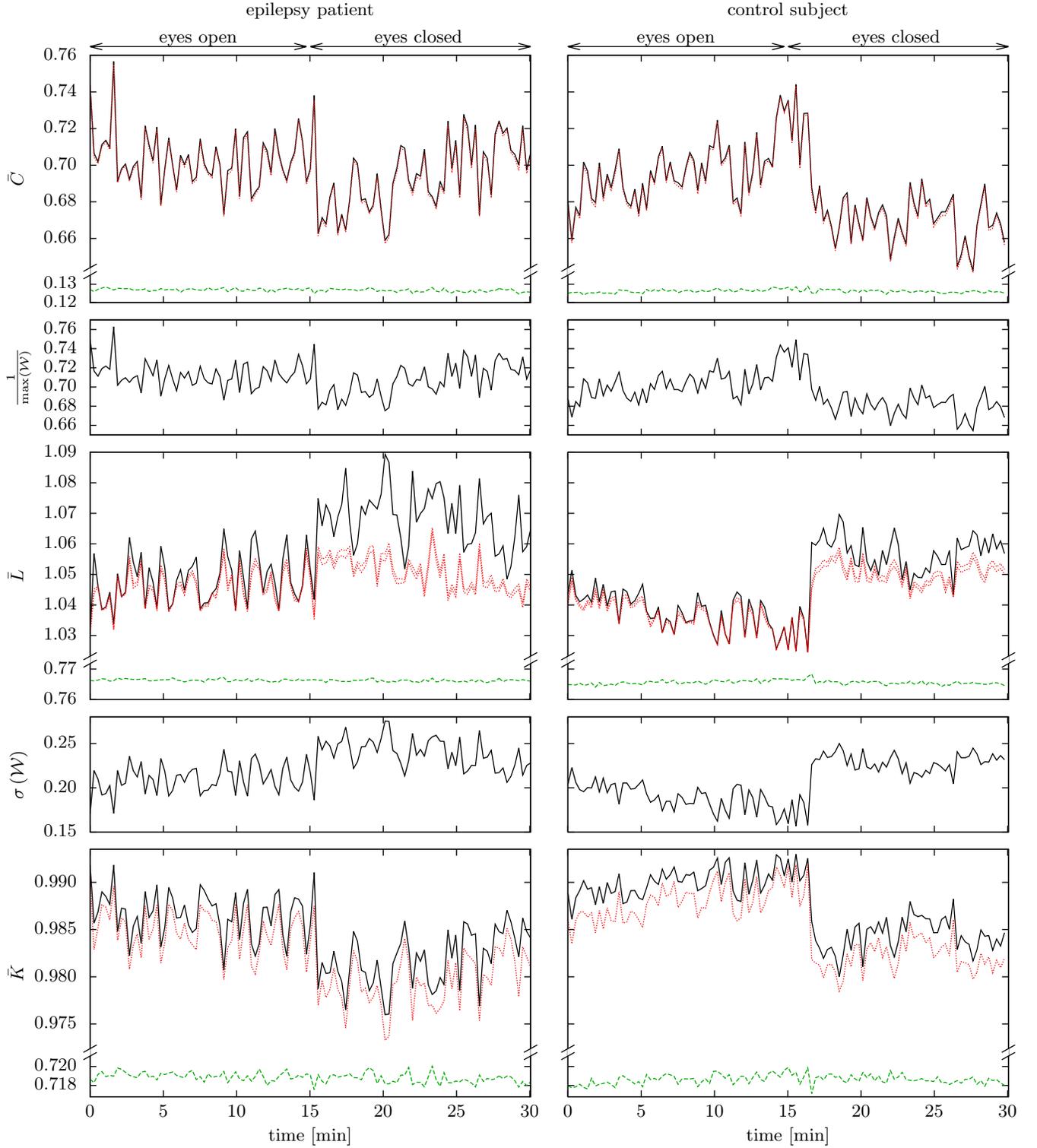}
	\caption{Temporal evolutions of clustering coefficients $\bar{C}$ (first row) and $\bar{K}$ (fifth row) and average shortest path length $\bar{L}$ (third row) of functional brain networks (black solid lines) and of weight-preserving surrogates (red dotted lines) and strength-preserving surrogates (green dashed lines) for these networks.
	For $\bar{L}$ we show the margins of standard deviation over $4096$ weight-preserving surrogates (red dotted lines).
	Standard deviations of $\bar{C}$ and $\bar{K}$ over the weight-preserving surrogates were too small to be displayed, the maximum standard deviation of $\bar{C}$, $\bar{K}$, and $\bar{L}$ over the $4096$ strength-preserving surrogates was $0.02$ each.
	For comparison, for the original networks we show the temporal evolutions of the inverse of the maximum weight $\frac{1}{\max\kl{\mathcal{W}}}$ (second row) and of the standard deviation $\sigma\kl{\mathcal{W}}$ of the edge weights (fourth row).}
	\label{fig:hirn}
\end{figure*}

EEG signals were split into consecutive non-overlapping segments of $4096$ data points (\unit[16.1]{s}) each.
For each segment we extracted the phases in a frequency-selective way using Morlet wavelets centred in the so-called alpha band (\unit[8--13]{Hz}) \cite{Niedermayer1993} and calculated the mean phase coherence $R_{ij}$ \cite{Mormann2000} as a measure for interdependence between signals recorded at sensors $i$ and $j$ (for simplicity's sake we omit the dependence on the segment in the following).
$R_{ij}$ is confined to the interval $\kle{0,1}$ where $R_{ij}=1$ indicates fully synchronised systems.
Network vertices were identified with sensors and edges between vertices $i$ and $j$ were assigned the weight $W_{ij} = R_{ij} - \bar{R} + 1$, where $\bar{R}$ is the average over all $R_{kl}$ with $k \neq l$.
For each of these networks, we generated $4096$ weight-preserving surrogates and $4096$ strength-preserving surrogates and calculated the clustering coefficients $\bar{C}$ and $\bar{K}$ as well as the average shortest path length $\bar{L}$ for the original and the surrogate networks.
Note, that for many applications, such as a test of a null hypothesis, fewer surrogates may suffice \cite{Schreiber2000a}.

In Fig.~\ref{fig:hirn} we show the temporal evolutions of $\bar{C}$, $\bar{K}$, and $\bar{L}$ for the functional networks of the epilepsy patient and the healthy control subject and for the corresponding weight- and the strength-preserving surrogates.
For both subjects we observed, on average, higher values of $\bar{L}$ and lower values of $\bar{C}$ and $\bar{K}$ during the eyes-closed condition.
There were, however, no clear-cut differences in $\bar{C}$ , $\bar{K}$, and $\bar{L}$ between the epilepsy patient and the control subject.
$\bar{L}$ during the eyes-open condition as well as $\bar{C}$ and $\bar{K}$ during the complete observation time were approximately equal for the original networks and the weight-preserving surrogates.
A property of the weight distribution $\mathcal{W}$, that we could identify as strongly correlated to $\bar{C}$, was the inverse of the maximum edge weight $\frac{1}{\max\kl{\mathcal{W}}}$.
We attribute this strong influence of $\max\kl{\mathcal{W}}$ mainly to its utilisation as a normalisation factor when calculating $\bar{C}$, since $\bar{K}$ did not exhibit such a strong correlation to $\frac{1}{\max\kl{\mathcal{W}}}$.
The temporal evolution of $\bar{L}$ was similar to that of the standard deviation of the edge weights of the original network $\sigma\kl{\mathcal{W}}$, while the temporal evolution of $\bar{K}$ was opposite to that of $\sigma\kl{\mathcal{W}}$.

Despite the mostly similar temporal evolutions of $\bar{C}$, $\bar{K}$, and $\bar{L}$ for the original and the weight-preserving surrogate networks, these characteristics always assumed higher values for the original networks than for any of the $4096$ surrogates.
Thus we can reject the null hypotheses $H_\mathcal{W}$, that the original networks are random under the constraint of their weight distribution $\mathcal{W}$.

When compared to the strength-preserving surrogates $\bar{C}$, $\bar{K}$, and $\bar{L}$ always assumed clearly higher values for the original networks, and we could not observe comparable temporal evolutions.
The null hypotheses $H_\mathcal{S}$, that the original networks are random under the constraint of their strength distribution $\mathcal{S}$, can be rejected as well.

Our findings indicate that the clustering coefficient $\bar{C}$ of the functional brain networks investigated here is predominantly determined by properties of the weight distribution $\mathcal{W}$.
Similar conclusions can be drawn for the clustering coefficient $\bar{K}$ and the average shortest path length $\bar{L}$, for the latter, however, for the eyes-open condition only.
In contrast, the clear differences between original and surrogate networks seen for $\bar{L}$ during the eyes-closed condition indicate that 
a considerable part of the value of this network-specific characteristic is not determined by the weight distribution $\mathcal{W}$ of the functional brain networks.
Whether these findings hold for all the data investigated in Ref.~\cite{Horstmann2010} needs further investigations, which will be published elsewhere.

\subsection{International Trade Networks}
As a second example we investigated the clustering coefficients $\bar{C}$ and $\bar{K}$ as well as the average shortest path length $\bar{L}$ of the International Trade Networks (ITN) \cite{Gleditsch2002,Garlaschelli2004,Saramaki2007,Fagiolo2009,Fagiolo2010,Bhattacharya2008} for the years 1948 to 2000.
The vertices of the ITNs are countries and the edge weights represent the amount of trade between the corresponding countries.
The number of vertices $n$ of the ITNs changes annually, growing from $n = 73$ in 1948 to $n = 187$ in 2000.
Since some binary properties of ITN of 1995 could be explained by a fitness model \cite{Garlaschelli2004}, it is conceivable that the structure of a weighted ITN is also governed by vertex-intrinsic parameters, which are reflected by the countries' total trade activity.
Since the latter corresponds to the vertex strengths, strength-preserving surrogates might detect such an influence.
As the number of vertices $n$ is preserved alongside with the strength distribution $\mathcal{S}$ and with the weight distribution $\mathcal{W}$, respectively, strength- or weight-preserving surrogates might help to detect a possible influence of this number on the network-specific characteristics.

To construct the networks from the data we followed Refs.~\cite{Saramaki2007, Bhattacharya2008} to determine the trade flow between two countries $i$ and $j$:
\[F_{ij} = \tfrac{1}{2}\kl{E_{ij}+I_{ij}+E_{ji}+I_{ji}}\]
where $E_{ij}$ and $I_{ij}$ denote the export and import from country $i$ to country $j$.
We determined the weights as $W_{ij} = F_{ij} / \bar{F}$, where $\bar{F}$ is the average over all $F_{ij}$ with $i \neq j$.
In each year we omitted countries, of which no trade was recorded at all
\footnote{For the year 1948 we also omitted the Koreas in order to obtain a connected network.}.
47\% of the edges of these networks were zero-weight edges.
47\% of this zero-weight edges were in turn to be attributed to missing data.
The latter (and probably some of the other zero-weight edges) are likely to correspond to small or negligible trade \cite{Gleditsch2002}.
For each year we calculated $\bar{C}$, $\bar{K}$, and $\bar{L}$ of the ITNs as well as of $4096$ weight-preserving surrogates and strength-preserving surrogates each.

\begin{figure*}[t]
	\input{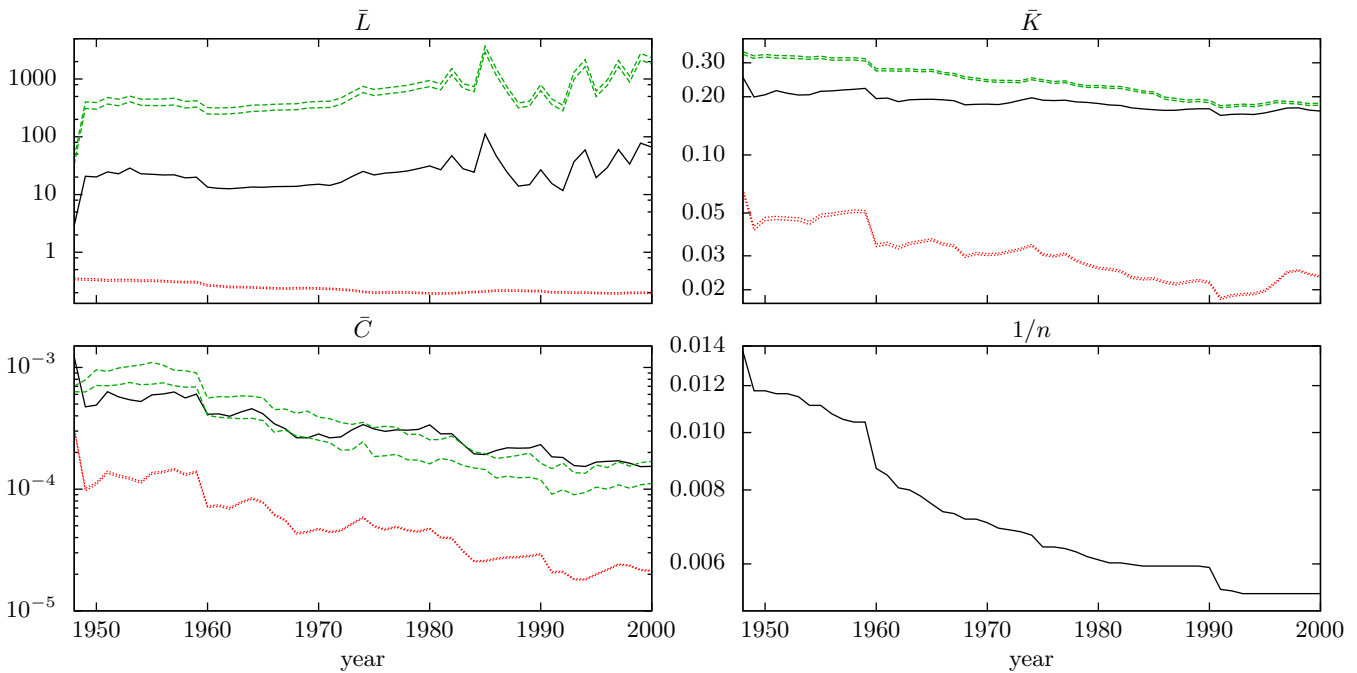}
	\caption{Top: temporal evolutions of clustering coefficient $\bar{K}$ (right) and average shortest path length $\bar{L}$ (left) of the International Trade Networks for the years 1948 to 2000 (black solid lines).
	Also shown are the margins of standard deviation for $4.096$ weight-preserving surrogates (red dotted lines) and $4.096$ strength-preserving surrogates (green dashed lines) for these networks.
	Bottom left: the same for the clustering coefficient $\bar{C}$.
	Bottom right: the inverse of the number of vertices $\frac{1}{n}$ for comparison.}
	\label{fig:wirtschaftsdaten}
\end{figure*}

In the top row of Fig.~\ref{fig:wirtschaftsdaten} we show the temporal evolutions of $\bar{K}$ and $\bar{L}$ for the ITNs and for the weight-preserving surrogates and strength-preserving surrogates.
For most years both characteristics of the ITNs clearly deviated from the respective values of the surrogates, and we thus can reject the null hypotheses $H_\mathcal{W}$ and $H_\mathcal{S}$, that the ITNs are random under the constraint of their weight distribution $\mathcal{W}$ or strength distribution $\mathcal{S}$, respectively.

We observed, however, considerable similarities in the temporal evolutions of $\bar{L}$ for the ITNs and for the strength-preserving surrogates, which approximately differed by a constant factor only (note, that the curves are almost parallel in the semi-logarithmic plot).
Hence it should be considered that the temporal changes of $\bar{L}$ can mainly be attributed to changes of $\mathcal{S}$ (i.e., of the annual relative trade volumes and the number of countries), though the absolute value of $\bar{L}$ cannot be attributed to them.
The similarities of the temporal evolutions of $\bar{K}$ between the ITNs and the surrogates are less dominant, but apparent for both types of surrogates.
This indicates that the temporal changes of $\bar{K}$ can only partially be attributed to changes of $\mathcal{W}$ or $\mathcal{S}$.
In the bottom right part of Fig.~\ref{fig:wirtschaftsdaten} we show the temporal evolution of $1/n$, which we observe to be similar to that of $\bar{K}$.
Increases of the number $n$ of countries, however, mostly coincide with separations of countries, which in turn may also affect $\mathcal{W}$ or $\mathcal{S}$.
Thus our findings do not resolve whether there is a direct influence of $n$ on $\bar{K}$.
The similarities of the temporal evolutions of $\bar{K}$ and $\bar{L}$ between the original networks and the surrogates indicate that there are only few changes in properties not to be attributed to the strength or weight distribution, respectively, and thus affirm that the ITNs' structure is mainly time-invariant \cite{Bhattacharya2008, Fagiolo2010}.

In the bottom left part of Fig.~\ref{fig:wirtschaftsdaten} we show the temporal evolutions of $\bar{C}$ for the ITNs and for the weight-preserving surrogates and strength-preserving surrogates.
We observe strong similarities in the temporal evolutions of $\bar{C}$ for the ITNs and the weight-preserving surrogates as well as of $\frac{1}{\max\kl{\mathcal{W}}}$ (not shown here).
These similarities and the fact that they are less pronounced for $\bar{K}$ affirm our findings in Sec.~\ref{hirn} that $\max\kl{\mathcal{W}}$ strongly influences $\bar{C}$ due to its use as a normalisation constant.

\section{Conclusions}
\label{concl}
We proposed a method to efficiently generate strength-preserving surrogates for complete weighted networks.
With strength-preserving surrogate networks and weight-preserving surrogate networks we reanalysis functional brain networks and investigated the International Trade Networks.
While we were examplarily regarding the clustering coefficient and the average shortest path length, surrogate networks can also be applied to investigate other network-specific characteristics.

For functional brain networks derived from an epilepsy patient and a healthy control subject during different states of vigilance we observed that the clustering coefficients $\bar{C}$ and $\bar{K}$ as well as the average shortest path length $\bar{L}$ are strongly dominated by properties of the weight distribution $\mathcal{W}$, namely, its standard deviation and its maximum.
Thus, previously reported differences between subjects as well as between states may be more easily identifiable by merely analysing properties of the distribution of interaction strengths $\mathcal{W}$.
Also, given the strong dependence of the clustering coefficient $\bar{C}$ on the maximum weight, other normalisations for $\bar{C}$ may be more appropriate for a comparison of networks.
It is even conceivable that, if the respective maximum weight of the networks under comparison is always held by the same edge, a comparison of the weights of this single edge suffices to identify differences.
In such a case a network approach to the data is questionable, since it is an overly complicated description of a simple aspect of the data.

For the International Trade Networks we observed that relative changes of the average shortest path length over the period 1948 to 2000 were reflected by the strength-preserving surrogates.
Similar results could also be obtained for the clustering coefficient $\bar{K}$, whose temporal evolution was also similar to that of the number of vertices.
This led us to assume that the relative changes were reflecting alterations of the vertex strengths, which are proportional to the trade volumes of the respective countries, or of the number of vertices.
Further investigations are necessary to clarify the impact of these influences on the ITNs' characteristics.

For both sets of empirical networks we could reject the null hypotheses corresponding to the applied surrogates in most cases.
This indicates that the networks are not only determined by their weight or strength distributions.
Our findings demonstrate that surrogate networks provide additional information about network-specific characteristics and thus can aid in their interpretation.

\section*{Acknowledgements}
We are grateful to Stephan Bialonski, Marie-Therese Kuhnert, and Alexander Rothkegel for helpful comments.	This work was supported by the Deutsche Forschungsgemeinschaft (Grant No.~LE660/4-2).

\appendix
\section{Mathematical Background}
\label{mathematics}
The $n$ linear equations, which correspond to the constraint of a given strength sequence of a (undirected, weighted, and complete) network are
\begin{equation}
S_i = \sum\limits_{j=1}^n W_{ij}.\label{eqn:LGS}
\end{equation}
Since there are $m \defi \frac{1}{2}n\kl{n-1}$ variables (the edge weights) in this system of linear equations, it has an $\kl{m-n}$-dimensional subspace of solutions, which we denote by $\Lambda$.
The set of non-negative solutions is the polytope $\Omega$.
For simplicity's sake we do not regard cases, in which the $\Lambda$-volume of $\Omega$ is $0$, e.g., if $S_i=0$ for any $i$ or if the network is star-shaped (i.e., there is one vertex, to which all non-zero-weight edges are adjacent).
With these omissions $\Omega$ is an $\kl{m-n}$-polytope and $\Lambda$ is its affine hull.

\subsection{Hit-and-Run Samplers}
The general procedure of a Hit-and-Run sampler for picking a random point or network, respectively, from $\Omega$ is \cite{Smith1984}
\begin{enumerate}
	\item Acquire some point $P^0\in\Omega$ and set the counter $h=1$.
	\item \label{pick} Pick a direction $D$ from the uniform distribution over a set of directions $\mathcal{D} \subset \mathbb{R}^m$.
	\item \label{transform} Pick a number $\zeta$ randomly from the uniform distribution on $\klg{\zeta \in \mathbb{R} \middle | P^{h-1} + \zeta D \in \Omega}$ and set $P^h = P^{h-1} + \zeta D$.
	\item If $h<t$, raise $h$ by $1$ and continue at \ref{pick}.
	Otherwise let $P^t$ be the random point.
\end{enumerate}
$P^t$ is approximately sampled from the uniform distribution, if $t$ is sufficiently large and if any two points of $\Omega$ are accessible from each other via some selected transformations as in step~\ref{transform} \cite{Smith1984}.

In our method, step~\ref{transform} of the Hit-and-Run procedure corresponds to a `tetragon transformation' (cf. Fig.~\ref{fig:Quadrate}) and each vector in $\mathcal{D}$ corresponds to a tetragon.
Such a vector has exactly four non-zero components, each of which has the same absolute value and corresponds to an edge of the tetragon.

\subsection{Accessibility of the polytope $\Omega$ by tetragon transformations}
In this section we show that any two points of $\Omega$ are accessible from each other via tetragon transformations, which is required for our Hit-and-Run sampler to sample uniformly from $\Omega$.
For this purpose we first show that there is a basis consisting only of vectors corresponding to tetragons (App.~\ref{basis}). From this follows that all vectors corresponding to tetragons form a spanning set of $\Lambda$ and thus each two points of the relative interior of $\Omega$ are accessible from each other.
Then we show that each point on the relative boundary of the polytope can be modified into one in the relative interior just with tetragon transformations (App.~\ref{ergo}) and vice versa.
Note, that despite this the probability, that any point on the relative boundary is sampled, is $0$.
The result is, however, important, if the original network is on the relative boundary.
Also, points in the relative interior near such an inaccessible point may only be accessible with a large number of transformations.

\subsubsection{A basis of $\Lambda$}
\label{basis}
Equation \ref{eqn:LGS} written as a matrix equation contains the following $n \times m$-matrix, if the variables (i.e., the edge weights) are ordered as described below (zeros are omitted):

\begin{center}
\noindent
\includegraphics{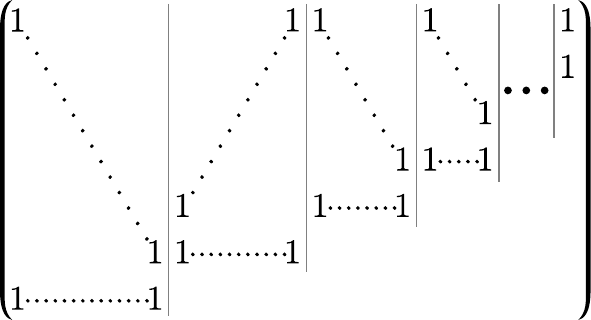}
\end{center}
The $i$-th row of this matrix corresponds to the (right-hand side of) equation $S_i = \sum_{j=1}^n W_{ij}$ and has an entry $1$ in all rows corresponding to weights $W_{ij}$ ($j \in \klg{1, \ldots, n} \backslash \klg{i}$).
Each column has exactly two non-zero entries, namely, the column corresponding to the edge weight $W_{ij}$ contains a $1$ in the rows $i$ and $j$.
The selected ordering of the weights may be separated into $n-1$ groups (as indicated by grey vertical lines), such that the $i$-th group contains the edges $W_{1,n-i+1}, \ldots, W_{n-i,n-i+1}$.
The second group's internal order is reversed to simplify the following conversions, which aim at determining a basis of $\Lambda$.

Subtracting all prior rows from the last one and then dividing the last row by $-2$ yields
\begin{center}
\noindent
\includegraphics{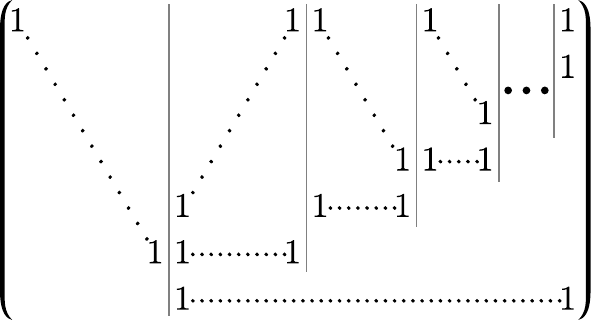}
\end{center}
Subtracting the last row from the two preceding rows yields
\begin{center}
\noindent
\includegraphics{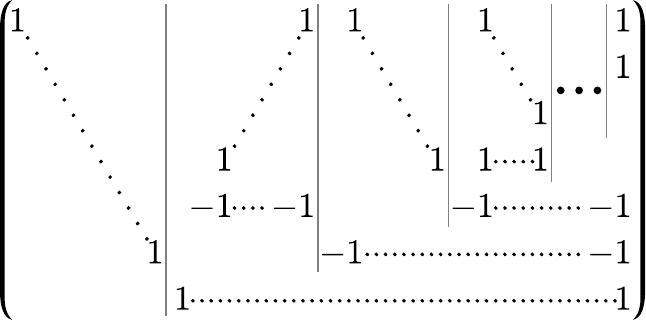}
\end{center}
Thus the column vectors of the following matrix are a basis of $\Lambda$:
\begin{center}
\noindent
\includegraphics{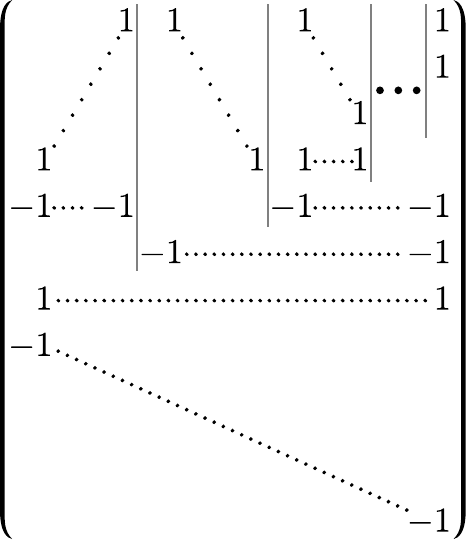}
\end{center}
The basis vectors of the first two groups contain exactly four non-zero components each.
Each basis vector in the remaining groups contains exactly six non-zero components and can be exchanged for a vector with four non-zero components by subtracting the vector in the first group that shares three non-zero components with it.
Thus there is a basis of $\Lambda$ only consisting of vectors with four non-zero components.
Since any of these vectors must solve $\forall i: 0 = \sum\limits_{j=1}^n W_{ij}$, the edges corresponding to its non-zero components must form a tetragon and thus all basis vectors correspond to a tetragon transformation.

\subsubsection{Accessibility of the relative boundary of the polytope $\Omega$ by tetragon transformations}
\label{ergo}
The points on the relative boundary of $\Omega$ are exactly those, which have at least one component that is zero.
Therefore, to show that any point on the relative boundary of $\Omega$ can be transformed into a point on the relative interior of $\Omega$ by tetragon transformations (and vice versa), it is sufficient to show that any zero-component (i.e., zero-weight edge) can be eliminated by tetragon transformations without creating a new one.

Let $W_{ij}$ be the zero-weight edge to be eliminated.
Since $S_i, S_j>0$, there must be at least one non-zero-weight edge adjacent to the vertices $i$ and $j$ (denoted by $W_{ik}$ and $W_{il}$, respectively).
\begin{enumerate}[I.]
	\item \label{normal} If $k \neq l$, the tetragon transformation that raises $W_{ij}$ and $W_{kl}$ by $\zeta = \tfrac{1}{2} \min \kl{W_{ik}, W_{il}}$ and lowers $W_{ik}$ and $W_{il}$ by the same amount eliminates the zero-weight edge $W_{ij}$ without creating a new one.
	\item If $k$ and $l$ can only be chosen such that $k = l$, there must be at least one non-zero-weight edge $W_{pq}$ with both $p$ and $q$ being unequal to both $k$ (otherwise the network would be star-shaped) and to either $i$ or $j$ (otherwise $n=3$).
	In this case first $W_{ip}$ or $W_{jp}$, respectively, and then $W_{ij}$ can be eliminated according to~\ref{normal}.
\end{enumerate}

\subsection{Comparison of tetragon transformations to other Hit-and-Run Samplers}
Standard choices for the direction set $\mathcal{D}$ are the unit sphere (Hypersphere Directions Hit-and-Run Sampler) or a basis (Coordinate Directions Hit-and-Run Sampler) \cite{Smith1996}\label{hypersphere}.
We expect our Hit-and-Run Sampler to be faster than the Hypersphere Directions Hit-and-Run Sampler, since the latter would require a transformation of each direction $D$ from a basis of $\Lambda$ to a basis of $\mathbb{R}^m$.
Moreover, all $m$ components need to be taken into account when choosing $\zeta$, while only four components need to be regarded during each tetragon transformation.
We also expect tetragon transformations to be more efficient than a Coordinate Directions Hit-and-Run Sampler, since they form a larger direction set $\mathcal{D}$ without increasing the computational burden per transformation.
Also for a Coordinate Directions Hit-and-Run Sampler the requirement of accessibility of all points may not be fulfilled.

\subsection{Extension to further constraints}
For some applications it may be desirable to generate surrogate networks that obey constraints further than the preservation of strengths or non-negative weights.
As long as tetragon transformations can transform every two points of the corresponding subset into each other, they may be used as direction set $\mathcal{D}$ for the Hit-and-Run-Sampler.
Otherwise or if in doubt, it can still be resorted to a Hypersphere Directions Hit-and-Run Sampler.
In the following we provide two examples, how further constraints can be incorporated into the Hit-and-Run sampler framework:
\begin{itemize}
	\item The constraint that the weights of the surrogates may not exceed a given maximum can be regarded analogously to the constraint of non-negative edge-weights.
	The set of possible surrogates is a smaller $\kl{m-n}$-polytope with $\Lambda$ as affine hull.
	Thus tetragon transformations can still transform all points of the relative interior of the new polytope into each other.
	App.~\ref{ergo} can be analogously applied to the constraint of a maximum weight.
	Problems may arise only in the case of zero-weight and maximum-weight edges in the same network.
	\item If the binary structure of the original network is to be preserved, zero-weight edges remain unaltered and the set of possible surrogates is a bounding sub-polytope of $\Omega$.
	For sparse networks, however, the requirement of accessibility of all points with tetragon transformations may not be fulfilled.
\end{itemize}

\section{Properties of the test statistics $\chi$}
\label{appprob}
If points $\mathcal{Q} = Q_1, \ldots, Q_a$ ($a \in \mathbb{N}$) are picked independently from the uniform distribution on $\Omega$, the probability $\pi$ that $c$ of them are in a given $\epsilon$-ball (or any other subset of $\Omega$) is binomially distributed:
\[\pi = B\kl{c, \varrho, a} \equiv \binom{a}{c} \varrho^a \kl{1-\varrho}^{a-c},\]
$\varrho \in \kle{0,1}$ being the fraction of $\Omega$'s volume that is occupied by the $\epsilon$-ball.
If a priori all $\varrho$ are equiprobable, the probability density of a given $\varrho$ is proportional to $\pi$.
If now $P_1, \ldots, P_a$ are also picked independently from the uniform distribution, the probability $p\kl{k}$ that $k$ of them are in the same $\epsilon$-ball is proportional to
\[\int\limits_0^1 B\kl{c,\varrho,a} B\kl{k, \varrho, a} d\varrho \ifed \hat{p}\kl{k}.\]
Multiple integrations by parts yield
\[\hat{p}\kl{k} = \frac{1}{2a+1} \binom{a}{c} \binom{a}{k} \binom{2a}{c+k}^{-1}\]
and normalisation finally results in
\[p\kl{k} = \hat{p}\kl{k} \kl{\sum\limits_{i=0}^a \hat{p}\kl{i}}^{-1} = \tilde{p}\kl{k} \kl{\sum\limits_{i=0}^a \tilde{p}\kl{i}}^{-1},\]
with $\tilde{p}\kl{k} \defi \binom{a}{k} \binom{2a}{c+k}^{-1}$.

For the calculation of $\chi$ several $\epsilon_i$-balls ($i \in \klg{0, \ldots b}$, $b \in \mathbb{N}$) around randomly picked points $R_1, \ldots, R_b$ are regarded, each containing exactly $c$ points from $\mathcal{Q}$.
The points $\mathcal{P} = P_1, \ldots, P_a$ were picked independently from an unknown distribution, and $k_i$ points from $\mathcal{P}$ are in the $\epsilon_i$-ball around $R_i$.
In this case, the higher $\hat{\chi} \defi \sum\limits_{i=1}^b p\kl{k_i}$ the more likely it is, that the points $\mathcal{P}$ are picked from the uniform distribution on $\Omega$.
Moreover, for $a,b \rightarrow \infty$ and $\frac{c}{a} \rightarrow 0$ ($\Rightarrow \epsilon_i \rightarrow 0 ~\forall i$) every local deviation from uniformity of $\mathcal{P}$'s distribution is captured and results in a decrease of $\hat{\chi}$.
Finally $\chi$ is obtained by normalizing $\hat{\chi}$ by its expected value in the case that $P_1, \ldots, P_a$ are picked independently from the uniform distribution:
\[\chi \defi \frac{ \sum\limits_{i=1}^b p\kl{k_i} } { b \sum\limits_{j=0}^a p\kl{j}^2 }
= \frac{ \sum\limits_{i=1}^{b} \tilde{p}\kl{k_i} \sum\limits_{j=0}^{a} \tilde{p}\kl{j} } {b  \sum\limits_{j=0}^a \tilde{p}\kl{j}^2 } .\]

\end{document}